\documentstyle[prl,multicol,aps]{revtex}

\input{epsf.tex}

\begin{document}
\title{ Nonlinear localized waves in a periodic medium }
\author{Andrey A. Sukhorukov and Yuri S. Kivshar} 
\address{Optical Sciences Centre, Australian National University, Canberra ACT 0200, Australia}
\maketitle
\begin{abstract}
We analyze the existence and stability of nonlinear localized waves in a periodic medium described by the Kronig-Penney model with a nonlinear defect. We demonstrate the existence of a novel type of stable nonlinear band-gap localized states, and also reveal an important physical mechanism of the oscillatory wave instabilities associated with the band-gap resonances.
\end{abstract}

\vspace*{-1 cm}

\begin{multicols}{2}
\narrowtext

Wave propagation in periodic media is associated with many interesting physical phenomena~\cite{yeh,book}. Modern technology allows to create different kinds of macro- and mesoscopic periodic and layered structures such as semiconductor superlattices and heterostructures, magnetic multilayers possessing the giant magnetoresistance, multiple-quantum-well structures, optical waveguide arrays, photonic band-gap materials and photonic crystal fibers, etc. The main feature of different periodic structures (which follows from the classical Floquet-Bloch theory) is the existence of forbidden frequency band gaps (or stop bands) where {\em linear waves undergo Bragg reflection} from the periodic structure~\cite{yeh}. However, many of the recently fabricated periodic structures exhibit pronounced nonlinear properties that give rise to qualitatively new physical effects such as {\em multistability} of a finite nonlinear periodic medium~\cite{multi} and energy localization in the form of {\em gap solitons}~\cite{gap}. Such effects are usually analyzed in the framework of the coupled-mode theory~\cite{coupled}, and they are associated with the nonlinearity-dependent tuning of the stop band as the wave intensity is increased.

Another fundamental reason for the wave localization is the presence of defects. As a matter of fact, artificially introduced inhomogeneities are often used as a powerful means of controlling wave scattering. However, the defect-induced localization in {\em nonlinear and periodic media} is largely an open area of research. We stress that the defect response is determined by the local field amplitude, and therefore the standard averaging procedure can not be directly applied to this kind of problems.

Recent experiments, e.g. the observation of optical gap solitons~\cite{gap_exp} and the control of coherent matter waves in optical lattices~\cite{kasevich}, as well as theoretical results such as the discovery of the oscillatory instability of gap solitons~\cite{gap_inst}, call for a systematic analysis of nonlinear effects in periodic structures and band-gap localized states {\em beyond the approximation provided by the coupled-mode theory}. Such an analysis is crucially important for determining {\em stability of nonlinear waves in periodic structures} because the wave instabilities can appear due to the mode coupling to other bands.

In this Letter, we present {\em the first analysis of the existence and stability of nonlinear localized waves in a periodic medium with multiple gaps in the transmission spectrum} that is valid beyond the coupled-mode theory. We consider a simple model where waves are localized in a layered medium by an intensity-dependent defect (or, in other words, they are guided by a thin-layer nonlinear waveguide). Assuming the applicability of our results to a variety of different physical systems (see below), we describe {\em four qualitatively different situations} and characterize, in the framework of a unified and systematic approach, the properties of {\em two types of nonlinear localized waves}~\cite{ref}. For the first time to our knowledge, we analyze {\em stability of nonlinear localized waves in a periodic medium} and reveal an important physical mechanism of wave instability associated with the band-gap resonances. We demonstrate also that several types of {\em stable band-gap localized states} can exist in the presence of nonlinearity.

{\em Model.} 
We describe localized waves in superlattices using the nonlinear Schr\"odinger equation for the wave envelope $\psi (x,t)$,
\begin{equation} \label{eq:nls}
     i \frac{\partial \psi}{\partial t} 
     + \frac{\partial^2 \psi}{\partial x^2}  
     + {\cal F}(I; x) \psi = 0,
\end{equation}
where $I \equiv |\psi|^2$ is the wave intensity, $t$ is time (or propagation variable), $x$ is the spatial coordinate, and the real function ${\cal F}(I; x)$ describes both {\em nonlinear} and {\em periodic} properties of the medium. We note that the system~(\ref{eq:nls}) is Hamiltonian, and for localized solutions the power $P = \int_{-\infty}^{+\infty} I(x)\; dx$ is conserved.

We seek stationary localized solutions of Eq.~(\ref{eq:nls}) in the form $\psi(x,t) = u(x) e^{i \omega t}$, where $\omega$ is the normalized frequency (or the propagation constant, in optics), and the real function $u(x)$ satisfies the equation:
\begin{equation} \label{eq:u0_inh}
  - \omega u + \frac{d^2 u}{d x^2} + {\cal F}(I; x) u = 0.
\end{equation}
We assume that the superlattice is linear, and nonlinearity appears only through the properties of an embedded localized defect. Then, if the corresponding width of the wave envelope is much larger than that of the defect, the inhomogeneity can be modeled by a delta-function and, in the simplest case, we can write 
   ${\cal F}(I; x) = \nu(x) + \delta(x) G(I)$, 
where the function $G(I)$ characterises the properties of the defect, and $\nu(x) \equiv \nu(x+h)$ describes an effective potential of the superlattice with the spatial period $h$. 

For such a local nonlinearity, the localized waves can be constructed with the help of a matching condition, by using the solutions of Eq.~(\ref{eq:u0_inh}) with ${\cal F}(I; x) = \nu(x)$, presented in the form of the Bloch-type wave functions~\cite{yeh}. 

{\em Band-gap structure and localized waves.}
If the effective periodic potential $\nu(x)$ is approximated by a piecewise-constant function (the so-called Kronig-Penney model), the solution on both sides of the defect can be decomposed into a pair of counter-propagating waves with the amplitudes $a(x)$ and $b(x)$,
\[
  u_b(x) = a(x) e^{- {\mu}(x) x} 
         + b(x) e^{{\mu}(x) x}, 
\]
where $\mu(x) = \sqrt{\omega - \nu(x)}$ is the local wavenumber.
As follows from the Floquet-Bloch theory, for a Bloch-wave solution the reflection coefficient $r(x) = b(x) / a(x)$ is a periodic function, i.e. $r(x) = r(x + h)$, and it is found from the following eigenvalue problem:
\begin{equation} \label{eq:eigen}
   T(x) \left( \begin{array}{l} 1 \\ r(x) \end{array} \right) 
    =
      \tau(x) 
      \left( \begin{array}{l} 1 \\ r(x) \end{array} \right) ,
\end{equation}
where $T(x)$ is a {\em transfer matrix} that describes a change of the wave amplitudes $\{a,b\}$ after one period $(x,x+h)$. An explicit expression for the transfer matrix can be found in Ref.~\cite{our_pre}, where it was also proven that ${\rm det}\;T \equiv 1$, and two linearly independent solutions of Eq.~(\ref{eq:eigen}) correspond to a pair of the eigenvalues $\tau$ and $\tau^{-1}$. Relation $\tau(\omega)$ determines a {\em band-gap structure} of the superlattice spectrum: the waves are {\em propagating}, if $|\tau| = 1$, and they are {\em localized}, if $|\tau| \ne 1$. In the latter case, a nonlinear defect can support {\em nonlinear localized waves}, and 
the wave amplitude at the defect is determined from the continuity condition~\cite{our_pre} at $x=0$ [$I_0 \equiv I(0)$]:
\begin{equation} \label{eq:zeta}
  G_0 \equiv G(I_0) = \zeta(\omega) .
\end{equation}
Here $\zeta = (\zeta^+ + \zeta^-)_{x=0}$, 
${\zeta^\pm} = \mu^\pm {\left( 1 - r^\pm \right)} 
                       {\left( 1 + r^\pm \right)}^{-1}$, 
and we denote with ``$+$'' and ``$-$'' the lattice characteristics on the right and the left side of the nonlinear defect, respectively, i.e. $\nu(x) = \nu^+(|x|)$, for $x>0$, and $\nu(x) = \nu^-(|x|)$, for $x<0$. Note that the reflection coefficients should correspond to the Bloch-wave solutions with the asymptotics $|u( x \rightarrow \pm \infty )| \rightarrow 0$. 

\begin{figure}
\setlength{\epsfxsize}{7.8cm}
\vspace*{-2mm}
\centerline{\mbox{\epsfbox{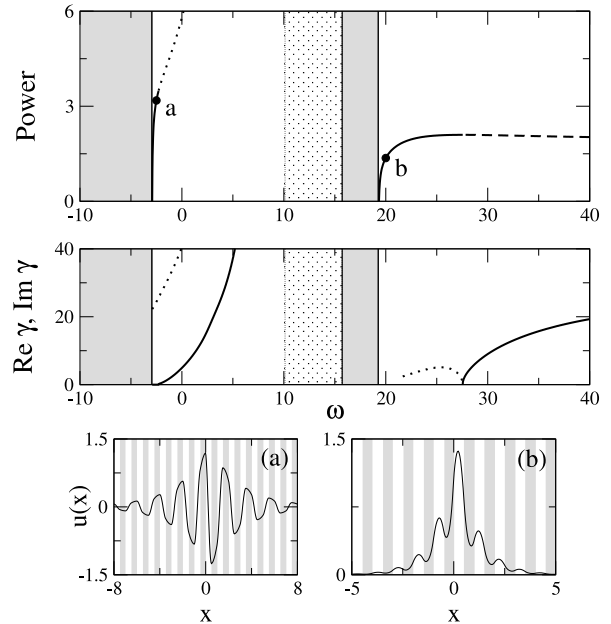}}}
\vspace*{1mm}
\caption{ \label{fig:pwr-sf-alp} \label{fig:bandgap}
Top: power vs. frequency dependences for the localized waves: solid~--- stable, dashed~--- unstable, and dotted~--- oscillatory unstable.
Middle: real (dotted) and imaginary (solid) parts of the eigenvalues associated with the wave instability.
Shading marks ``waveguiding'' (white) and ``anti-waveguiding'' (dotted) localization regimes inside the band gaps.
Bottom:~the localized states corresponding to the marked points (a,b) in the top plot; shading marks the areas with smaller $\nu$.
The lattice parameters are $h=1$, $\nu(x)=0$ for $n-1/2 < x/h < n$, and $\nu(x)=30$ for $n < x/h < n+1/2$, where $n$ is integer.
The impurity characteristics are $\alpha = 0.5$, $\beta=1$.}
\end{figure}

Relation~(\ref{eq:zeta}) allows us to identify different types of nonlinear localized states. We notice that such localized states, supported by an attractive nonlinear defect ($G_0 > 0$), exist in the so-called {\em waveguiding} regime, when $\zeta(\omega) > 0$. Additionally, localization can occur at a repulsive nonlinear defect ($G_0 < 0$) in the {\em anti-waveguiding} regime, provided $\zeta(\omega) < 0$. A characteristic band-gap structure for a two-component superlattice is presented in Fig.~\ref{fig:bandgap} (top). Using the guided-wave terminology, we notice that the first (semi-infinite) band-gap corresponds to the conditions of {\em the total internal reflection} (IR) and, therefore, its dispersion properties should be similar to those of the conventional waveguiding regime. Localized waves in this band resemble conventional solitary waves modulated by a periodic structure [see Fig.~\ref{fig:bandgap}(b)]. In contrast, both waveguiding and anti-waveguiding regimes can occur at smaller $\omega$, where {\em band-gaps} appear due to the resonant Bragg-type reflection (BR) from the periodic structure, so that the localized waves are somewhat similar to the gap solitons composed of mutually coupled backward and forward propagating waves [see Fig.~\ref{fig:bandgap}(a)].

To study {\em the linear stability}, we consider the evolution of small-amplitude perturbations of the localized state presenting the solution in the form
\[
 \psi (x,t) 
 = \left\{ u(x) + v(x) e^{i \gamma t} 
                + w^{\ast}(x) e^{-i \gamma^{\ast} t} 
   \right\} e^{i \omega t} ,
\]
and obtain the linear eigenvalue problem for $v(x)$ and $w(x)$. Then, from the condition of mode localization we define the so-called {\em Evans function},
$Y( \gamma  ) = 
   \left[ G_1 - \zeta(\omega+\gamma) \right]
   \left[ G_1 - \zeta(\omega-\gamma) \right]
   - \left( G_1 - G_0 \right)^2 $, 
where $G_1 \equiv G_0 + I_0 G^{\prime}(I_0)$, the intensity $I_0$ is calculated for an unperturbed solution, and the prime stands for the derivative. Eigenvalues are found as zeros of the Evans function, $Y( \gamma ) = 0$, and 
the corresponding eigenmode solutions fall into one of the following categories:
(i)~{\em internal modes} with real eigenvalues describe periodic oscillations (``breathing'') of the localized state, (ii)~{\em instability modes} correspond to purely imaginary eigenvalues, and (iii)~{\em oscillatory instabilities} can occur when the eigenvalues are complex.

\begin{figure}
\setlength{\epsfxsize}{7cm}
\centerline{\mbox{\epsfbox{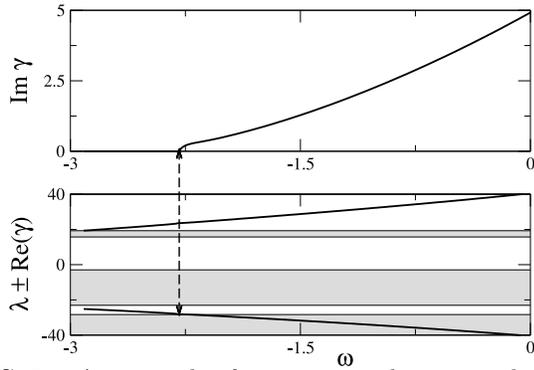}}}
\caption{ \label{fig:imode-sf-alp}
An example of a resonance that occurs between an internal mode of the localized wave and a bang-gap edge and leads to an oscillatory instability.}
\end{figure}

To demonstrate the basic stability results, we consider a localized defect possessing a cubic nonlinear response,
  $G(I) = \alpha + \beta I$.
Under proper scaling the absolute value of the nonlinear coefficient~$\beta$ can be normalized to unity, so that $\beta=+1$ corresponds to {\em self-focusing} and $\beta=-1$ to {\em self-defocusing} nonlinearity. Localization depends also on the sign of the linear coefficient~$\alpha$, which defines the defect response at small intensities: {\em attractive} if $\alpha>0$, and {\em repulsive}, otherwise. Therefore, below we consider {\em four qualitatively different examples} that correspond to the different signs of~$\alpha$ and~$\beta$.

\begin{figure}
\setlength{\epsfxsize}{7.8cm}
\vspace*{-2mm}
\centerline{\mbox{\epsfbox{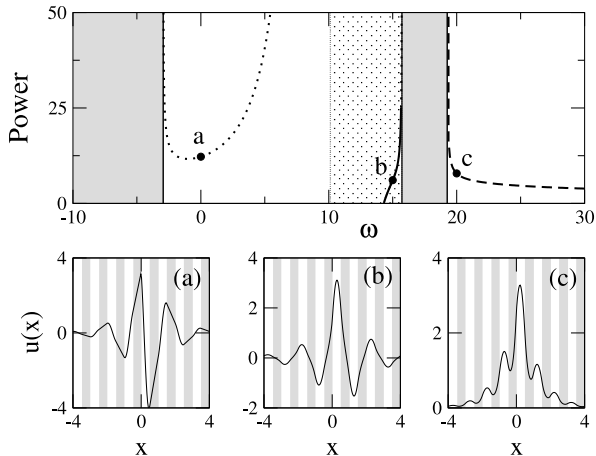}}}
\vspace*{1mm}
\caption{ \label{fig:pwr-sf-alm}
Power vs. frequency, and (a-c)~wave profiles for $\alpha = -5$, $\beta = 1$. Notations are the same as in Fig.~\ref{fig:pwr-sf-alp}.}
\end{figure}

{\em Self-focusing nonlinearity.}
First we consider the properties of the localized waves supported by a defect with a self-focusing nonlinearity ($\beta=+1$), and attractive linear response ($\alpha>0$). Such waves already exist in the linear limit at the frequencies $\omega_b$ defined by the equality $\zeta(\omega_b)=\alpha$, and they correspond to the waveguiding regime only (white regions in Fig.~\ref{fig:pwr-sf-alp}, top). Since an IR wave is a fundamental eigenstate of the self-induced waveguide, it can be demonstrated that the conditions of the Vakhitov-Kolokolov (VK) stability theorem~\cite{VK} are satisfied, and the IR states are {\em unstable} if and only if $d P / d \omega < 0$. The critical point $d P / d \omega = 0$ corresponds to a ``collision'' between two internal modes of the localized wave at the origin, as illustrated in Fig.~\ref{fig:pwr-sf-alp} (middle). 
For the BR states, the VK criterion provides only a necessary condition for stability, since the higher-order localized states can also exhibit oscillatory instabilities. Indeed, we notice that in the linear limit 
there always exists an internal mode corresponding to a resonant coupling between the BR and IR band-gaps, since $Y \left( \omega_b^{\rm (IR)}-\omega_b^{\rm (BR)} \right) = 0$. We perform extensive numerical calculations and find that this mode leads to an oscillatory instability of BR waves when the value $(\omega-{\rm Re}\;\gamma)$ moves outside the band gap; it happens when the intensity exceeds a threshold value (see Figs.~\ref{fig:pwr-sf-alp} and~\ref{fig:imode-sf-alp}).

\begin{figure}
\setlength{\epsfxsize}{7.8cm}
\vspace*{-2mm}
\centerline{\mbox{\epsfbox{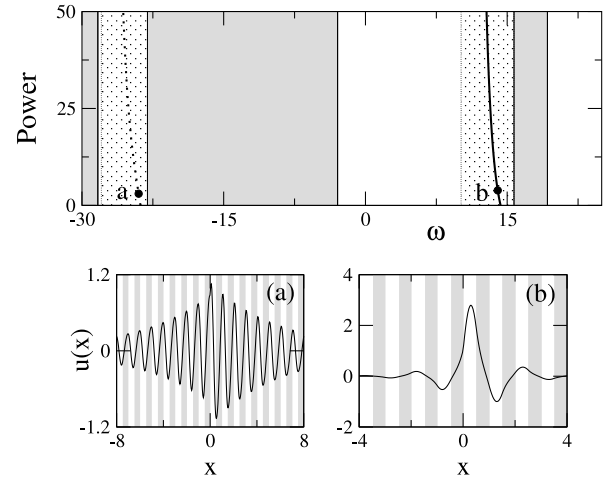}}}
\vspace*{1mm}
\caption{ \label{fig:pwr-df-alm}
Power vs. frequency, and (a,b)~wave profiles for $\alpha = -5$, $\beta = -1$.
Notations are the same as in Fig.~\ref{fig:pwr-sf-alp}.}
\end{figure}

Next, we study the linearly repulsive defect ($\alpha<0$) with  self-focusing nonlinearity. In this case, the waveguiding-type states bifurcate from the band-gap edges [with $G_0 = \zeta(\omega_b) = 0$] and the stability criteria are the same as in the case $\alpha>0$ discussed above, see Fig.~\ref{fig:pwr-sf-alm}. The major difference from the previous case is the existence of {\em the anti-waveguiding states} at small intensities, $I_0 < |\alpha/\beta|$. We find that such waves {\em are stable in the lowest photonic band gap} [such as one shown in Fig.~\ref{fig:pwr-sf-alm}(b)], while in higher band gaps they exhibit oscillatory instabilities due to a resonance with the higher-frequency states. Different types of the localized waves and their stability are summarized in Fig.~\ref{fig:pwr-sf-alm}.

{\em Self-defocusing nonlinearity.}
Let us now consider the case $\alpha<0$ and $\beta=-1$, when the total response of the defect is negative for any intensity, i.e. $G(I) < 0$. In this case, the localized waves can only exist in the {\em anti-waveguiding} regime (see Fig.~\ref{fig:pwr-df-alm}, top), the nonlinear waves continue the linear impurity states, and their frequency decreases at higher powers. We find that the localized states corresponding to the lowest BR gap, such as that shown in Fig.~\ref{fig:pwr-df-alm}(b), are always {\em stable}. However, in other band gaps the localized states [e.g., the one shown in Fig.~\ref{fig:pwr-df-alm}(a)] can exhibit {\em oscillatory instability} due to the physical mechanism identified in other cases~--- {\em a resonance with the lower band gaps}.

\begin{figure}
\setlength{\epsfxsize}{7.8cm}
\vspace*{-2mm}
\centerline{\mbox{\epsfbox{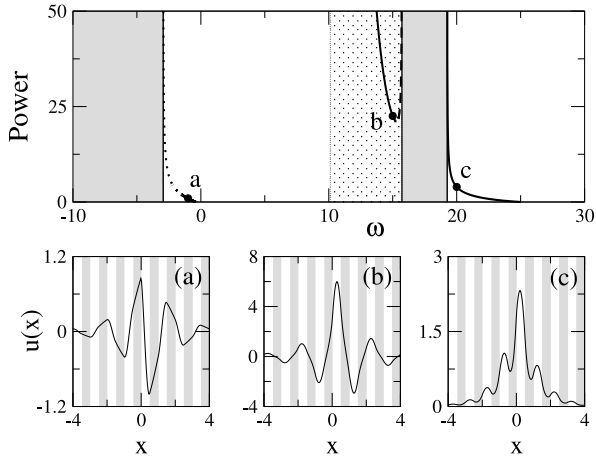}}}
\vspace*{1mm}
\caption{ \label{fig:pwr-df-alp}
Power vs. frequency, and (a-c)~wave profiles for $\alpha = 5$, $\beta = -1$. Notations are the same as in Fig.~\ref{fig:pwr-sf-alp}.}
\end{figure}

Finally, we study the case of a linearly attractive defect ($\alpha>0$) possessing a self-defocusing nonlinearity ($\beta=-1$). At small intensities, the defect is attractive, and it can support localized waves in a {\em waveguiding} regime, in both IR and BR gaps, as shown in Fig.~\ref{fig:pwr-df-alp}. Performing the stability analysis, we find that IR waves are always {\em stable}, while oscillatory instabilities appear for higher band-gap states. Since at higher intensities, i.e. for $I_0 > |\alpha/\beta|$, the defect response changes its sign, a new type of localized waves can exist in the {\em anti-waveguiding} dispersion region, bifurcating from the band-gap edge. It is possible to demonstrate that for such waves {\em the VK stability criterion becomes inverted}, i.e. the waves are unstable if  $d P / d \omega > 0$, as shown in Fig.~\ref{fig:pwr-df-alp} (top). This happens because the signs of both the nonlinear response and effective dispersion are altered compared to the IR waves supported by a self-focusing defect. All higher band-gap states can exhibit oscillatory instabilities.

{\em Concluding remarks.}
We have analyzed the existence and stability of nonlinear localized waves in one-dimensional periodic structures considering the simplest nonlinear generalization of the Kronig-Penney model with a nonlinear impurity for which both the nonlinear wave classification and stability analysis can be carried out in {\em a complete and systematic way}. Taking into account many common features of nonlinear guided waves and impurity modes in stratified and disordered media, on one side, and the self-trapped states and solitary waves in homogeneous nonlinear media, on the other side, we expect that many of the results described above will be found in more realistic physical models of nonlinear periodic media. Such cases include the electron self-trapping and locking states in cuprates and semiconductor superlattices~\cite{a1}; nonlinear guided waves in optical superlattices~\cite{a2}; impurity modes in photonic band-gap materials~\cite{a3}, magneto-optical periodic structures~\cite{a4}, and photonic crystal fibers~\cite{pcf}; coherent matter waves in optical lattices~\cite{a5}, etc. In particular, the stability of nonlinear localized impurity modes is a crucial issue for creating tunable band-gap materials where gaps could be controlled by changing the input light intensity.

We are indebted to O.~Bang and C.~M.~Soukoulis for useful collaboration.
The work has been partially supported by the Planning and Performance Fund of the Institute of Advanced Studies.

\end{multicols}

\begin{references}
\bibitem{ref}
Using the optics analogy, we notice that such two types of guided waves correspond to the conventional waveguiding, due to the total internal reflection, and to the band-gap states, due to the Bragg-type reflection.
\end{references}
\end{document}